\documentclass[a4paper,11pt]{article}
\pdfoutput=1 

\usepackage{jcappub} 
\usepackage[T1]{fontenc} 

\usepackage{natbib}
\usepackage{cancel}
\usepackage{enumitem}    
\usepackage{amsthm}         
\usepackage{amsmath} 	   
\usepackage{amssymb}       
\usepackage{amsfonts}
\usepackage{graphicx} 
\usepackage{dblfloatfix}	    
\usepackage{verbatim}
\usepackage{epstopdf}
\usepackage{tensor}
\usepackage{epsfig,multicol,bbm}
\usepackage{color}
\usepackage{colortbl}
\usepackage{float}
\usepackage{multirow}
\usepackage{array}

\usepackage{subfig}

\usepackage{afterpage}

\allowdisplaybreaks

\title{Gauge/frame invariant variables for the numerical relativity study of cosmological spacetimes}
\author[a]{Anna Ijjas}

\affiliation[a]{Center for Cosmology and Particle Physics, Department of Physics, New York University, New York, NY, 10003, USA}

\emailAdd{ijjas@nyu.edu}

\abstract{To numerically evolve the full Einstein equations (or modifications thereof), simulations of cosmological spacetimes must rely on a particular formulation of the field equations combined with a specific gauge/frame choice. Yet truly physical results cannot depend on the given formulation or gauge/frame choice.  In this paper, we present  a resolution of the gauge problem and, as an example, numerically implement it to evaluate our  previous work  on contracting spacetimes.}


\begin{document}
\maketitle 
\raggedbottom

\section{Introduction}

In recent years, there has been an increased interest in developing non-perturbative numerical relativity analyses of cosmological spacetimes (see {\it e.g.}, \cite{Ijjas:2022qsv}).  Such studies are relative newcomers in the field of cosmology that, for the past fifty plus years, has been dominated by perturbative/field theory methods. They nevertheless have a great potential to significantly impact the field because these type of analyses enable us to answer a variety of key cosmological questions that could not be addressed using earlier field theory methods. The cosmological initial conditions problem or the non-linear evolution of gravitational theories beyond Einstein are immediately obvious examples but they only represent the first few of many opportunities for discovery. 

In this paper, to further aid the emergence of this new field, we present a set of diagnostic tools to evaluate the results of numerical relativity studies of cosmological spacetimes in a way that does not depend on the specific choice of formulation, coordinate gauge or frame. We illustrate the power and usefulness of the tool kit by applying it to several examples from existing studies testing the robustness of early-universe models to initial conditions.

A reason why non-perturbative numerical relativity studies of cosmological spacetimes started to appear only a few years ago was the lack of existing computational infrastructure. 
It took several decades to meet the many complex mathematical and computational challenges until the first example of a successful numerical relativity simulation of a strongly gravitating system -- a binary black hole merger \cite{Pretorius:2006tp} -- was obtained. These challenges involved, on the one hand, finding a proper formal dressing of the field equations (choice of appropriate formulation, gauge and frame) \cite{Shibata:1995we,Baumgarte:1998te,Garfinkle:2001ni} as well as developing numerical techniques (such as constraint damping \cite{Gundlach:2005eh} ) to stably {\it evolve} the full non-linear system of Einstein field equations. On the other hand, one had to find {\it diagnostic tools} to reliably extract the relevant physics from the simulations \cite{NewmanPenrose}.

Numerical relativity analyses of cosmological spacetimes greatly benefit from the tools and methods developed to evolve binary black hole spacetimes and other compact object mergers. Yet naturally, this infrastructure does not translate trivially to numerically evolving strongly gravitating cosmological spacetimes and must first be adapted   to appropriately capture the specific physics context. So far all relevant studies have been focused on finding proper evolution schemes for cosmological spacetimes; see {\it e.g.}, \cite{Ijjas:2022qsv}. An evolution scheme is always associated with a particular formulation of the field equations combined with a specific gauge/frame choice. That means, the dynamical variables of a non-perturbative, numerical relativity simulation are always variables expressed in a particular gauge/frame. To extract the physics,  the dynamical variables that are being evolved in the simulation have to be combined into gauge/frame invariant quantities upon concluding the numerical computation. 

For example, to establish that a mechanism is a robust smoother that drives spacetime towards a spatially flat, homogeneous and isotropic Friedmann-Robertson-Walker (FRW) state, it is essential that the mechanism is probed for a sufficiently general set of initial conditions including those that lie outside the perturbative regime of FRW spacetimes in {\it any} frame/gauge. 
Short of a coordinate independent diagnostics, inhomogeneities or anisotropies in one gauge used for the underlying numerical relativity simulation could turn out to  be coordinate artifacts, arising purely from the particular gauge choice. In this case, robustness to initial conditions is not  established, even if the mechanism appears to result in a smooth spacetime. For this reason, to arrive at a reliable result, it is indispensable to complement existing and future work by a gauge/frame independent diagnostics.

Here, we present a way of resolving the gauge problem utilizing the Weyl tensor and associated invariants known as the Weyl curvature and Chern-Pontryagin invariant. 
These quantities enable us to unambiguously characterize inhomogeneities and anisotropies as well as the absence thereof in non-perturbative, numerical relativity simulations of cosmological spacetime in a gauge/frame invariant way and without losing information from the gauge/frame dependent formulation underlying the simulation. 
We start by introducing the basic formalism in Sec.~\ref{sec:formalism}. The bulk of this paper involves numerically implementing this formalism and exemplifying its usefulness and power by applying it to the case of contracting spacetimes in Secs.~\ref{sec:num}-\ref{sec:results}. 
In companion papers \cite{Garfinkle:2023vzf, Ijjas:2023dnb}, we employ our formalism to evaluate existing work studying the robustness of slowly contracting and inflating spacetimes to initial conditions.

\section{Formalism}
\label{sec:formalism}

In non-perturbative, numerical relativity calculations, one solves a coupled non-linear system of partial differential equations (PDEs) that involves a particular choice of formulation, gauge and/or frame of the underlying field equations. 
A ‘formulation’ (sometimes called an initial value formulation) of a given theory is the representation of the underlying
system of differential equations obtained by choosing a particular coordinate system. Note, though, that a formulation is not
equivalent to a gauge choice. Different gauge conditions can be implemented in the same formulation. For example, most calculations is cosmology use the Arnowitt-Deser-Misner (ADM) form \cite{Arnowitt:1959ah}, which is a coordinate based (3+1) formulation of the field equations. Common ADM gauge choices for perturbed spacetimes are unitary or Newtonian. Here, we shall employ a tetrad formulation \cite{Estabrook:1964zk,Buchman:2003sq,Ijjas:2020dws} and combine it with a particular frame and coordinate gauge choice that we detail in Appendix~\ref{app-dynvar}. Yet another formulation, which is heavily used in mathematical and numerical relativity, is the generalized harmonic form, see {\it e.g.}, \cite{Garfinkle:2001ni}. In principle, any gauge can be implemented in this formulation by choosing appropriate harmonic source functions. 

Naively, one would expect that the choice of formulation or gauge  is a matter of taste or convenience due to local Lorentz invariance. In reality, though, most formulations, gauge or frame choices do not lend themselves to numerical evolution. Rather, only a small subset of possible formulations, gauge or frame choices meets certain mathematical and computational criteria that are necessary for stable numerical evolution of the field equations; for an introductory review, see {\it e.g.}, \cite{Ijjas:2020dws}. For example, the Einstein field equations in harmonic form yield a strongly hyperbolic system that is a`well-posed' formulation of the initial value problem, meaning that the PDE system admits a unique solution that continuously depends on the initial conditions. 
However, the ADM form combined with algebraic gauge conditions for the lapse function and the shift vector (as commonly used in cosmology \cite{Bardeen:1980kt, Bardeen:1983qw}) is ill-suited for numerical evolution because it yields a weakly hyperbolic system that is ill-posed \cite{Ijjas:2018cdm}. 

To interpret specific results of a numerical relativity calculation -- that is, to extract observables or compare with other simulations -- we need diagnostic tools  that translate the gauge/frame dependent output of the computation into gauge/frame invariant quantities.  This problem has already been addressed in the context of simulating black hole and other compact object mergers. In the black hole case, the dynamical variables are combined into components of the Weyl tensor in the Newman-Penrose form \cite{NewmanPenrose}, yielding a precise and accurate gauge invariant description of the outgoing gravitational radiation  in the far-field regime, on asymptotically-flat spacetimes.

As we will detail below, in numerical relativity simulations of cosmological spacetimes,  the Weyl tensor is also useful for characterizing the results of the simulations in both a gauge and frame independent way. However, we do not simply borrow the diagnostic quantities used in the context of compact object mergers, as suggested by, {\it e.g.}, Ref.~\cite{Macpherson:2018btl,Munoz:2022duf}. Instead, we use tools that are specifically aimed at the needs of non-perturbative cosmological analyses, that involve (but are not limited to) characterizing the genericity of initial conditions or understanding basics of the non-linear evolution towards or away from flat FRW spacetimes.

\subsection{The Weyl tensor}

The trace-free part of the Riemann curvature tensor $\tensor{R}{_\mu_\nu_\sigma_\rho}$ is called the {\it conformal Weyl curvature tensor} (henceforth referred to as {\it Weyl tensor}),
\begin{equation}
\label{Weyl-coo}
\tensor{\cal C}{_\mu_\nu_\rho_\sigma}
\equiv \tensor{R}{_\mu_\nu_\rho_\sigma}
+ {\textstyle \frac12}\left( \tensor{g}{_\mu_\nu_\rho_\zeta}\tensor{R}{^\zeta_\sigma} 
+ \tensor{g}{_\mu_\nu_\zeta_\sigma}\tensor{R}{^\zeta_\rho}\right)
 + {\textstyle \frac16}R \tensor{g}{_\mu_\nu_\rho_\sigma},
\end{equation} 
where $g_{\mu\nu}$ denotes the spacetime metric, $\tensor{R}{_\mu_\nu}\equiv \tensor{R}{^\sigma_\mu_\sigma_\nu}$ denotes the Ricci curvature tensor, $R\equiv \tensor{R}{^\mu_\mu}$ denotes the Ricci curvature scalar 
and $\tensor{g}{_\mu_\nu_\rho_\sigma} \equiv  g_{\mu\rho}g_{\nu\sigma} - g_{\mu\sigma}g_{\nu\rho}$; see \cite{Weyl:1918pdp}.
Throughout, spacetime indices $(0-3)$ are Greek and spatial indices $(1-3)$ are Latin.  

The Weyl tensor  describes the `non-local' part of the gravitational field, {\it i.e.}, inhomogeneities and anisotropies of the spacetime geometry that are not sourced by a local stress-energy source. As its name suggests, $\tensor{\cal C}{_\mu_\nu_\rho_\sigma}$ is invariant under conformal transformations of the metric. 

Note that, in spacetime dimensions $D\leq3$, the Weyl tensor identically vanishes but in 3+1 dimensional spacetimes, the Weyl tensor is generally non-zero. In fact, $\tensor{\cal C}{_\mu_\nu_\rho_\sigma}\equiv 0$ if and only if spacetime is conformally-flat, {\it i.e.}, 
\begin{equation}
g_{\mu\nu} = \psi(t,{\bf x}) \eta_{\mu\nu},
\end{equation}
where $\psi(t,{\bf x})$ is the conformal factor and $\eta_{\mu\nu}={\rm diag}(-1,1,1,1)$ denotes the (3+1) dimensional flat (Minkowski) metric. 

It is customary to introduce the (left) dual of the Weyl tensor,
\begin{equation}
\label{def-dualWeyl}
{}^{\ast}\tensor{\cal C}{_\mu_\nu_\rho_\sigma}
\equiv \frac12 \tensor{\chi}{_\mu_\nu^\tau^\zeta}\tensor{\cal C}{_\tau_\zeta_\rho_\sigma},
\end{equation}
with $ \tensor{\chi}{_\mu_\nu_\tau_\zeta}\equiv - \sqrt{|-g|}\tensor{\varepsilon}{_\mu_\nu_\tau_\zeta}$ being the totally anti-symmetric Levi-Civita 4-form and $\tensor{\varepsilon}{_\mu_\nu_\tau_\zeta}$ being the Levi-Civita tensor.

The Weyl tensor and its dual can be combined to yield two curvature invariants,
\begin{equation}
\label{def-weylscalar}
{\cal C} \equiv \tensor{\cal C}{^\mu^\nu^\rho^\sigma}\tensor{\cal C}{_\mu_\nu_\rho_\sigma},
\end{equation}
called the {\it Weyl curvature}, and 
\begin{equation}
\label{def-CPscalar}
{\cal P} \equiv  {}^{\ast}\tensor{\cal C}{^\mu^\nu^\rho^\sigma}\tensor{\cal C}{_\mu_\nu_\rho_\sigma},
\end{equation}
called the {\it Chern-Pontryagin invariant}.

As we will demonstrate in the remainder of this paper, the Weyl tensor and the associated invariants ${\cal C}$ and ${\cal P}$ enable us to unambiguously characterize inhomogeneities and anisotropies as well as the absence thereof in non-perturbative, numerical relativity simulations of cosmological spacetime in a gauge/frame invariant way and without losing information from the gauge/frame dependent formulation underlying the simulation. 

\subsection{Worked example}
\label{sec:diagnostics}

In a series of papers \cite{Cook:2020oaj,Ijjas:2020dws,Ijjas:2021gkf,Ijjas:2021wml,Ijjas:2021zyf,Kist:2022mew}, we performed extensive non-perturbative, numerical relativity simulations of contracting spacetimes.
To exemplify the power and usefulness of the Weyl tensor in the context of cosmological spacetimes, we will implement the gauge/frame independent diagnostics into our existing numerical scheme, as detailed below. 
Our existing work naturally lends itself for this purpose since it involves the study of evolving cosmological space-times with initial conditions outside the perturbative regime of flat FRW geometries. In addition, we have gained great control over what type of initial data leads to complete smoothing (flat FRW) and what (rare) initial conditions and/or potentials lead to partial smoothing or the lack thereof. That is, we can test our proposed diagnostic tools under a variety of conditions and for a variety of cosmological spacetimes, all under complete theoretical and numerical control.
In a companion paper \cite{Garfinkle:2023vzf}, we employ our formalism to evaluate existing work studying the robustness of inflating spacetimes to initial conditions.

To evolve contracting spacetimes, we numerically solve the full (3+1) dimensional Einstein-scalar field equations,
\begin{eqnarray}
\label{E-eq1}
R_{\mu\nu} - \frac12 g_{\mu\nu}R &=& \nabla_{\mu}\phi\nabla_{\nu}\phi -  g_{\mu\nu}\left( \frac12\nabla_{\lambda}\phi\nabla^{\lambda}\phi +V(\phi) \right)\\
\label{E-eq2}
\Box \phi &=& V_{,\phi}
\end{eqnarray}
 in mean curvature normalized, orthonormal tetrad form; for details of the scheme, see Appendix~\ref{app-dynvar}. Here, $\Box=\nabla^{\lambda}\nabla_{\lambda}$ denotes the covariant d'Alembertian.
 The stress-energy is being sourced by a canonical scalar field $\phi$ that is minimally-coupled to Einstein gravity and has a negative exponential potential  
\begin{equation}
\label{def-potential}
V(\phi)=-V_0\exp(-\phi/M),
\end{equation}
where the constant pre-factor $V_0>0$ denotes the potential energy density at $\phi=0$ and $M>0$ denotes the characteristic mass scale of $\phi$.  In the examples presented below, $V_0=0.1$.
Throughout, we express the scalar field $\phi$ in reduced Planck units, $M_{\rm Pl} =8\pi G_{\rm N} = 1$, with $G_{\rm N}$ being Newton's constant; $V_0$ in units of $M_{\rm Pl}^2(K_0/3)^2$, where $K_0$ is the initial mean curvature; and we express length and time in units of $K_0$.


In formulations that involve a 3+1 split, it proves useful to project the Weyl tensor and its dual introduced in Eqs.~\eqref{Weyl-coo} and \eqref{def-dualWeyl} onto the timelike congruence ${\bf e}_0$:
\begin{eqnarray}
\label{def-elec}
E_{\alpha\beta} &\equiv&  {\cal C}_{\alpha\mu\beta\nu} \tensor{e}{_0^\mu}\tensor{e}{_0^\nu},
\\
\label{def-magn}
H_{\alpha\beta} &\equiv&  {}^{\ast}{\cal C}_{\alpha\mu\beta\nu} \tensor{e}{_0^\mu}\tensor{e}{_0^\nu}.
\end{eqnarray}  
Due to their behavior at short wavelengths/high frequencies which resembles the electric and magnetic fields in Maxwell's theory,  
$E_{\alpha\beta}$ and $H_{\alpha\beta}$ are called the `electric' and `magnetic' part of the Weyl tensor, respectively \cite{Weyl:1918pdp,matte_1953}. Throughout, the beginning of the alphabet ($\alpha, \beta, \gamma$ or $a,b,c$) denotes tetrad indices and the middle of the alphabet ($\mu, \nu, \rho$ or $i,j,k$) denotes coordinate indices.

With the variables and gauge conditions of our scheme as defined in Appendix.~\ref{app-dynvar},  the electric and magnetic components of the Weyl tensor take the following form:
\begin{eqnarray}
\label{def-elec}
E_{ab}  &=&
{\Sigma}_{ab} 
- \tensor{{\Sigma}}{_{\langle a}^c} \tensor{{\Sigma}}{_{b\rangle c}} 
+ {}^3{R}_{\langle ab\rangle} - {\textstyle \frac12}{\pi}_{ab}, 
\\
\label{def-magn}
H_{ab} &=& 
{\textstyle \frac12}\tensor{n}{_c^c}{\Sigma}_{ab}
- 3\tensor{\Sigma}{^c_{(a}}\tensor{n}{_{b) c}} 
+ \tensor{n}{_c_d}\Sigma^{cd}\delta_{ab}
+ \tensor{\epsilon}{^c^d_{(a}} \left( \tensor{E}{_{c}^i} \partial_i  - A_c \right)  \tensor{\Sigma}{_{b)d}},
\end{eqnarray} 
where the 3-curvature term ${}^3{R}_{\langle ab\rangle}$ denotes the trace-free Ricci-tensor of the 3-metric induced on the spacelike hypersurfaces orthogonal to the timelike congruence ${\bf e}_0$ and is given by
\begin{equation}
{}^3{R}_{\langle ab\rangle} = 
\tensor{E}{_{\langle a}^i} \partial_i \tensor{A}{_{b\rangle} }
+ 2\tensor{n}{_{\langle a}^c}\tensor{n}{_{b\rangle c}} - \tensor{n}{_c^c}n_{\langle ab\rangle}
- \tensor{\epsilon}{^c^d_{\langle a}}\Big( \tensor{E}{_{c}^i} \partial_i  - 2A_c \Big) \tensor{n}{_{b\rangle d}},
\end{equation}
with angle brackets denoting traceless symmetrization,  {\it i.e.}, $X_{\langle ab\rangle}\equiv X_{(ab)}-{\textstyle \frac13}\tensor{X}{_c^c}\delta_{ab}$. 
For a canonical scalar, the anisotropic stress $\pi_{ab}$ takes the form
\begin{equation}
\pi_{ab} = S_{\langle a}S_{b\rangle} 
.
\end{equation}
We verified that our expressions agree with earlier derivations (before frame or gauge fixing), see, {\it e.g.} Ref.~\cite{vanElst:1996dr}.

Note that, by construction, both $E_{\alpha\beta}$ and $H_{\alpha\beta}$ are symmetric and trace-free, and both tensors are orthogonal to the timelike congruence, meaning that only their spatial components, $E_{ab}$ and $H_{ab}$ are non-zero. The two spatial 3-tensors are symmetric and trace-free with each of them having 5 independent components. The 10 independent components of $E_{ab}$ and $H_{ab}$ fully characterize the 10 independent components of the Weyl tensor.

It is straightforward to verify that we can re-express the two curvature invariants ${\cal C}$ and ${\cal P}$ defined in Eqs.~(\ref{def-weylscalar}-\ref{def-CPscalar}) by only using the  spatial 3-tensors $E_{ab}$ and $H_{ab}$. Substituting into Eqs.~(\ref{def-weylscalar}-\ref{def-CPscalar}), we obtain the simple expressions:
\begin{eqnarray}
\label{Weyl-invariant}
\label{def2-weyl}
{\cal C} &=& 8\Big( E_{ab}E^{ab} - H_{ab}H^{ab}\Big),\\
\label{def2-cp}
{\cal P} &=& 16 E_{ab}H^{ab}.
\end{eqnarray}

In the relativity literature, it has been common to introduce additional invariants in order to characterize various exact solutions by their symmetry properties ({\it e.g}, Petrov classification) \cite{Petrov:2000bs}, or to describe the outgoing gravitational radiation on asymptotically-flat spacetimes  (Newman-Penrose formalism). 
However,  these additions to ${\cal C}$ and ${\cal P}$  are more than what is needed when it comes to using mathematical and numerical relativity to addressing key issues in modern cosmology.

Mathematical and numerical relativity provide a unique contribution  in that they  provide tools to study mechanisms that drive  spacetimes, in particular, our Hubble patch towards or away from the perturbative regime of flat FRW geometries. For example, these tools can help address the cosmological initial conditions problem or study the stability of theories beyond Einstein as well as their observational predictions.  None of these issues can be addressed using the conventional methods of cosmological perturbation theory and effective field theory.
As we will exemplify below, following the evolution of the two invariants ${\cal C}$ and ${\cal P}$ relative to the mean curvature complemented by the eigenvalue evolution of the electric and magnetic components at representative spatial points provide a complete set of diagnostic tools that is sufficiently precise and powerful in evaluating corresponding simulations of cosmological spacetimes in a frame/gauge independent way. 

\section{Numerical implementation}
\label{sec:num}

As in our previous studies, we numerically evolve the Einstein-scalar field equations~(\ref{eq-E-ai-H}-\ref{eq-S-a-H}) after specifying our initial data set such that it satisfies the constraint equations~(\ref{hamiltonian-const}-\ref{const-phi-H}). In the following, we give a brief summary of our methods, for details, see, {\it e.g.}, Refs.~\cite{Ijjas:2020dws,Ijjas:2021gkf}.

\subsection{Initial data}

\label{sec:init_cond}



As is common in numerical relativity studies, to specify our initial data, we adapt the York method \cite{York:1971hw} and define the spatial metric of the initial $t_0$-hypersurface to be conformally-flat,  
\begin{align}
    g_{ij}(t_0,\vec{x}) = \psi^4(t_0,\vec{x})\;\delta_{ij}, 
\label{eq:conf-flat}
\end{align}
where $\psi(t_0,\vec{x})$ the conformal factor.  
Together with the mean curvature $\Theta_0^{-1}$ at $t_0$, which we can freely set,  $g_{ij}(t_0)$ fixes the components of the spatial curvature tensor,
\begin{align}
    \bar{n}_{ab}(t_0,\vec{x}) &= 0, 
 \\
    \bar{A}_b(t_0,\vec{x}) &= -2\psi^{-1}(t_0,\vec{x})\;{\bar{E}{}_b}^i(t_0,\vec{x})\;\partial_i\psi(t_0,\vec{x}),
\label{eq:a0}
\end{align}
as well as the tetrad vector components,
\begin{align}
    {\bar{E}{}_a}^i(t_0,\vec{x}) = \psi^{-2}(t_0,\vec{x})\;\Theta_0^{-1}\;{\delta_a}^i.
\end{align}
Note that choosing a conformally-flat initial metric is not a true restriction on the initial data. Rather, it is a straightforward way to ensure the constraints are satisfied. As we will see in the examples presented below, unlike constraint satisfaction, which is propagated by the Einstein-scalar equations, conformal flatness is broken within only a few integration steps.

A great advantage that comes with employing the York method is the freedom to  semi-analytically define the initial scalar field distribution,
\begin{align}
\phi(t_0,\vec{x}) &= f_x\cos(n_x x+h_x) 
+ \phi_0,
\label{eq:phi0} 
\end{align}
its conformally-rescaled velocity, 
\begin{align}
\bar{Q}(t_0,\vec{x}) &\equiv \psi^6(t_0,\vec{x})\bar{W}(t_0,\vec{x}) 
= \Theta_0\times\Big(q_x\cos(m_x x+d_x) 
+ Q_0\Big),
\label{eq:q0}
\end{align}
as well as the divergence-free part of the conformally-rescaled anisotropy (or shear) tensor, 
\begin{align}
    \bar{Z}_{a b}^{0} (t_0,\vec{x}) \equiv \psi^6(t_0,\vec{x})\bar{\Sigma}_{ab}^0(t_0,\vec{x})
= \Theta_0\times
\begin{pmatrix}
b_{2} && \xi &&0 \\
\xi && b_{1}+a_{1} \cos x && a_{2} \cos x \\
0 &{\;}&  a_{2} \cos x &{\;}& -b_{1}-b_{2}-a_{1} \cos x
    \end{pmatrix}.
\label{eq:z_ab0}
\end{align}
The components $\bar{Q}$ and $\bar{Z}_{ab}$ of the conformally rescaled scalar field velocity and shear tensor are specified through the parameters $\phi_0, Q_0, f_x, q_x, n_x, m_x$, $h_x, d_x$, $a_1, a_2, b_1, b_2$ and $\xi$ which we freely and independently set for each simulation run. 

The sinusoidal form of the spatial variations reflects that the boundary conditions are periodic, $0\leq x \leq 2\pi$, with $0$ and $2\pi$ identified. Here, we only show a single mode for the shear and two modes for the field's velocity; these can be replaced by a sum of fourier modes with different amplitudes, wavenumbers and phases.


The initial data is then completed by numerically computing the conformal factor, $\psi(t_0,\vec{x})$, and the rest of the shear tensor, $\bar{Z}_{ab} - \bar{Z}^0_{ab}$, using the Hamiltonian and momentum constraints, respectively.

\subsection{Evolution}

We evolve the hyperbolic-elliptic system~(\ref{lapse-eq-H}, \ref{eq-E-ai-H}-\ref{eq-S-a-H}) using standard finite difference methods, as detailed in Refs.~\cite{Cook:2020oaj,Ijjas:2020dws,Ijjas:2021gkf,Ijjas:2021wml}: We discretize the system using second-order accurate spatial derivatives and a three-step method for time integration employing the Iterated Crank-Nicolson algorithm.  At each sub-step, we first solve the elliptic equation~\eqref{lapse-eq-H} 
using the tridiagonal matrix algorithm and then update the hyperbolic equations~(\ref{eq-E-ai-H}-\ref{eq-S-a-H})
to the next Iterated Crank-Nicolson sub-step. In the simulations presented below, we use a grid of 1024 spatial points with $\Delta x = 2\pi/1024$ and timesteps $\Delta t = -0.5\Delta x$. 

In addition, after updating the hyperbolic equations, we compute at each time step all components of the Weyl tensor $E_{ab}, H_{ab}$ given by Eqs.~(\ref{def-elec}-\ref{def-magn}) as well as the Weyl curvature ${\cal C}$ given by Eq.~\eqref{def2-weyl}, the Chern-Pontryagin invariant ${\cal P}$ given by Eq.~\eqref{def2-cp} and all  eigenvalues associated with the spatial 3-tensors $\bar{\Sigma}_{ab}, \bar{n}_{ab}, E_{ab}, H_{ab}$ at select sample points, for details, see Appendix~\ref{sec:appendixEV}. 

To demonstrate the convergence of our code, the error and convergence were analyzed for a broad range of examples using the same methods as detailed in the Appendices of Refs.~\cite{Ijjas:2020dws,Ijjas:2021gkf}. Our code shows no
signs of numerical instability and exhibits clear second order convergence at early times. At later times when a smooth, ultralocal spacetime develops, we observe that the convergence improves to third order.

We do {\it not} recommend using our diagnostic tools in combination with recent trends  (see, {\it e.g.}, Ref.~\cite{Munoz:2022duf}) which `refurbish' existing evolution schemes not developed for the needs of numerical relativity studies of cosmological spacetimes.
There are several pitfalls when refurbishing existing schemes originally developed to simulate, {\it e.g.}, binary black hole or compact object mergers. To name only a few examples, issues can arise from limitations on the choice of initial data \cite{East:2015ggf}, or from limitations on how long the numerical integration can be  run before encountering bad behavior ({\it e.g.}, due to stiffness or formation of singularities \cite{Corman:2022alv}), or from the inappropriate choice of matter form ({\it e.g.}, fluid as used for neutron start simulations \cite{Macpherson:2018btl}). All of these choices can significantly impact the quality of the studies as well as the validity of conclusions. The diagnostics presented here cannot heal these issues and are only valid in the context of schemes that are first shown to avoid these problems. 

\section{Results}
\label{sec:results}

In this Section, we present a representative example of slowly contracting spacetimes. In this example, the scalar field has a moderately sloped negative exponential potential as defined in Eq.~\eqref{def-potential} with $M=0.2$. The  free data $\{\Theta_0, \bar{Z}^0_{ab}, \phi, \bar{Q}\}$ as introduced in Sec.~\ref{sec:init_cond}  are specified through the following parameters:
\begin{align}
\label{init1-1}
&\Theta_0 = 3.00;\\
&a_1=0.50,\; a_2=0.50,\; b_1=1.80,\; b_2=-0.15,\; \xi=0.01;
\\
& f_x = 0.00,\; n_x=1,\; h_x =-1.00,\; \phi_0=0.00;\\
\label{init1-4}
& q_x =0.51,\; m_x=2,\; d_x=-1.70,\; Q_0 = 0.10.
\end{align}   
The characteristic feature of this example is the tiny, downhill directed average initial velocity $Q_0=0.1$ corresponding to an initial kinetic energy density ${\textstyle \frac12}\bar{W}^2 \sim {\cal O}(0.01)$ (in units of the mean curvature).
The combination of a negative exponential potential and slightly downhill directed initial average field velocity is the key to all of spacetime being smoothed within 40 $e$-folds of contraction of the inverse mean curvature. That is, all of spacetime rapidly converges to the flat FRW geometry and we observe the Hubble radius undergoing 100 plus $e$-folds of contraction while physical distances shrink only by a factor of 3.

This example has been previously studied in Ref.~\cite{Ijjas:2020dws} but the analysis was done using gauge/frame dependent quantities. Working with an existing example, our goal is to illustrate the power and utility of the frame/gauge independent diagnostic tools presented above in Sec.~\ref{sec:formalism} to earlier methods.

In Ref.~\cite{Ijjas:2020dws}, we have shown that this is one of the `borderline' cases of slow contraction, which is precisely the reason why we chose to consider it here. The example is borderline in the sense that the combination of moderately sloped exponential potential and tiny initial kinetic energy lead to a comparatively long phase of ${\cal O}(10)$ $e$-folds of contraction in the inverse mean curvature $\Theta$ before complete smoothing is reached. (NB: In general, complete smoothing is reached within only a few $e$-folds.) The fact that reaching complete smoothing takes longer enables us to observe the different phases of the evolution in great detail.

\subsection{Evolution of the curvature invariants}

\begin{figure}[!tb]%
    \centering
\includegraphics[width=0.8\linewidth]{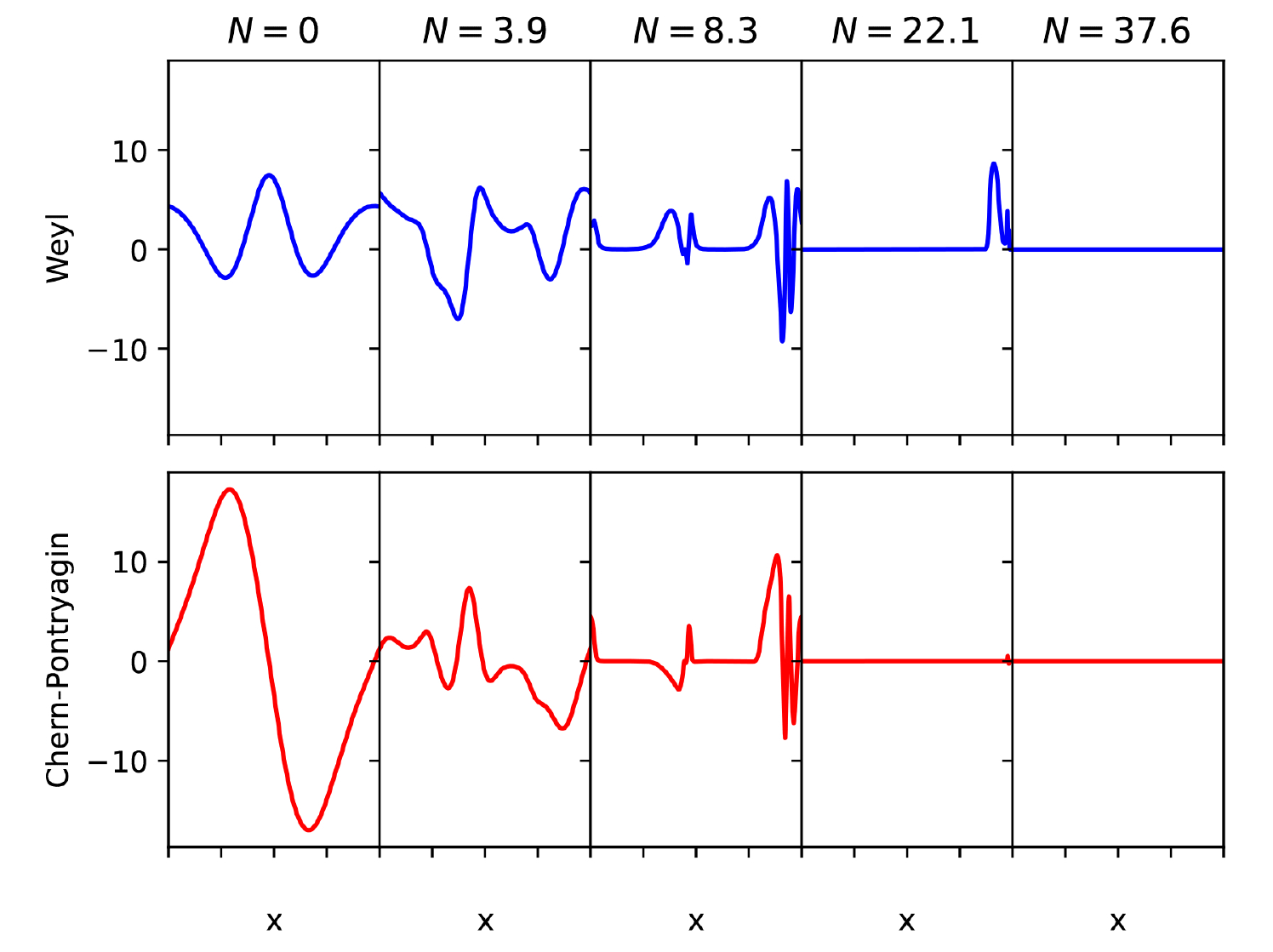}%
    \caption{Snapshots of the Weyl curvature $\bar{\cal C}$ (upper panel) and Chern-Pontryagin invariant  $\bar{\cal P}$ (lower panel) at sample times $N$ during slow contraction corresponding to $M=0.2$ and the initial data set given in Eqs.~(\ref{init1-1}-\ref{init1-4}). The $x$ axis represents the entire simulation domain and the $y$ axis has dimensionless units by construction of the rescaled invariants. The variable $N$ denotes the number of $e$-folds of contraction in the inverse mean curvature $\Theta$. Complete smoothing ($\bar{\cal C}, \bar{\cal P}\lesssim {\cal O}(10^{-10})$) is reached everywhere by $N\sim38$. }%
    \label{fig:smg-init}%
\end{figure}
In Figure~\ref{fig:smg-init}, we present snapshots of the Weyl curvature scalar and the Chern-Pontryagin invariant in Eqs.~(\ref{def2-weyl}-\ref{def2-cp}) relative to the mean curvature (which ratios we mark with a bar): 
\begin{eqnarray}
\label{def2-weyl-H}
\bar{\cal C} &=& 8\Big( \bar{E}_{ab}\bar{E}^{ab} - \bar{H}_{ab}\bar{H}^{ab}\Big),\\
\label{def2-cp-H}
\bar{\cal P} &=& 16 \bar{E}_{ab}\bar{H}^{ab},
\end{eqnarray}
where $\bar{E}_{ab}\equiv E_{ab}/\Theta^{-2}$ and $\bar{H}_{ab}\equiv H_{ab}/\Theta^{-2}$. Evaluating the curvature invariants relative to the mean curvature ({\it i.e.}, in the homogeneous limit, the Hubble radius)  yields the observationally relevant quantities in the exact same way as the density parameters $\Omega_i \equiv \rho_i/(3\Theta^{-2})$ (see, {\it e.g.}, Eqs.~4.1-4.3 in Ref.~\cite{Ijjas:2020dws}) yield the observationally relevant quantities as opposed to the mere densities $\rho_i$. 
  
Manifestly, the initial data we consider lie far outside the perturbative regime of flat FRW spacetimes as indicated by the fact that both $\bar{\cal C}(x)$ and $\bar{\cal P}(x)$ reach values of ${\cal O}(10)$. From the next snapshot at $N\simeq4$, it is clear that the evolution is strongly non-linear and the initial symmetry of $\bar{\cal C}(x)$ and $\bar{\cal P}(x)$ at $N=0$ is not preserved. This is indicative of the fact that the initial data chosen is sufficiently general and not close to a stationary point. By $N\simeq8$, the first spacetime points start to settle in the flat FRW state ($\bar{\cal C}, \bar{\cal P}\lesssim {\cal O}(10^{-10})$) but most of spacetime is still in an inhomogeneous, spatially curved and anisotropic state.

Importantly, numerical relativity studies are purely classical evolutions that do not include quantum perturbations, {\it e.g.}, to ${}^3\bar{R}$. For cosmological applications, though, to qualify as sufficiently homogenizing, isotropizing and flattening the universe ({\it i.e.}, driving spacetime sufficiently close to `flat FRW'),  a mechanism must suppress $\bar{\cal C} \sim {\cal O}( ({}^3\bar{R})^2)$ and $\bar{\cal P} \sim {\cal O}( ({}^3\bar{R})^2)$ to be much less than the quantum contributions to $({}^3\bar{R})^2$, which are known to be ${\cal O}(10^{-10})$ based on observations of the cosmic microwave background.
(Strictly speaking, $\bar{\cal C}$ and $\bar{\cal P}$ approaching zero are only necessary conditions for a geometry to be {\it conformally equivalent} to flat FRW. However, if the (scalar field) matter is minimally coupled in the Einstein frame with $\Omega_{m}\simeq1$ and $|\Omega_k|+|\Omega_{\Sigma}|\lesssim{\cal O}(10^{-5})$, as in all examples considered here, then $\bar{\cal C}$ and $\bar{\cal P}$ approaching zero guarantees flat FRW specifically.)

By $N\simeq22$, though, with the exception of a small (proper) volume, all of spacetime is smoothed to the required level. Complete smoothing is reached by $N\simeq38$. We ran the simulation beyond $N=120$ $e$-folds and observed the system to remain in the flat FRW state with no indication of any instability. The curvature invariants continued to shrink and reached values less than $10^{-40}$ by the end of the simulation.

\begin{figure}[!tb]%
    \centering
\includegraphics[width=0.75\linewidth]{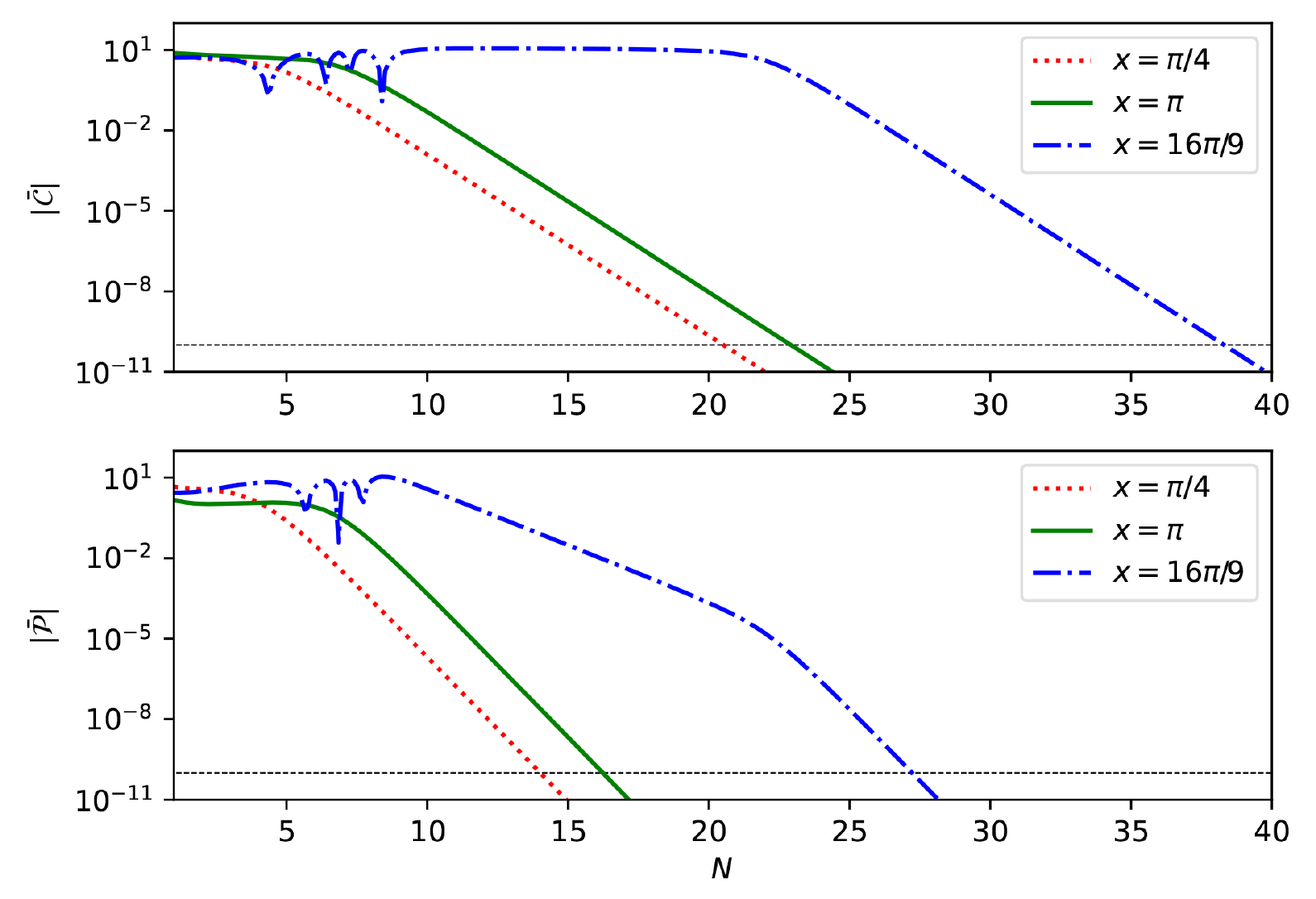}%
    \caption{(Log-scale) Evolution of the Weyl curvature $\bar{\cal C}(x)$ (upper image) and Chern-Pontryagin invariant $\bar{\cal P}(x)$ (lower image) at select spatial points: $x=\pi/4$ (red dotted line),  $x=\pi$ (green solid line) and  $x=16\pi/9$ (blue dash-dotted line) as a function of $e$-fold time $N$ corresponding to the case in Fig.~\ref{fig:smg-init}. The black dashed line marks the (observationally motivated) upper bound on $\bar{\cal C}$ and $\bar{\cal P}$ for the flat FRW state.}%
    \label{fig:smg-numpts_log}%
\end{figure}
The fact that different spacetime points reach the stable flat FRW attractor state at different times  is reflected in Fig.~\ref{fig:smg-numpts_log} where we depict the evolution of  $\bar{\cal C}$ and $\bar{\cal P}$  at sample spatial points in different spacetime regions. 
Note that we have chosen the three points to be outside of each others' `Hubble radius' (as measured by the mean curvature $\Theta$) at $N=0$ and remain outside each others' Hubble radius for the entire simulation.

Each of the three points starts out in a strongly curved, inhomogeneous and anisotropic state with both curvature invariants having values of ${\cal O}(10)$ but the three points undergo different evolutions. As indicated in Fig.~\ref{fig:smg-init}, the first spacetime point $x=\pi/4$ is in a region that first settles (red dotted line) in the flat FRW state. The second point $x=\pi$ reaches (green solid line) the flat FRW state at $N\simeq 22$, only a few $e$-folds later than the first point. The third point $x=16\pi/9$ is in the small (proper volume) region that reaches the flat FRW state at $N\simeq38$. 

We note that,  even though spacetime points undergo somewhat different evolution histories and reach the flat FRW attractor state at different times, we observe two main common features: (1) for all three points, the Chern-Pontryagin invariant decreases before the Weyl curvature decreases; (2) both curvature invariants   decrease exponentially. Intriguingly, the exponential decrease starts for both invariants at the same $e$-fold time and coincides with the time when the spacetime point reaches the ultralocal state, {\it i.e.}, when the gradient terms in the respective evolution equations become negligible; see Ref.~\cite{Ijjas:2021gkf,Ijjas:2023dnb}. 

These features are signatures of ultralocality. Although 
ultralocality is commonly presented as a frame/gauge dependent feature, these results show that it has a frame and gauge independent interpretation. 

\begin{figure}[!tb]%
    \centering
\includegraphics[width=0.8\linewidth]{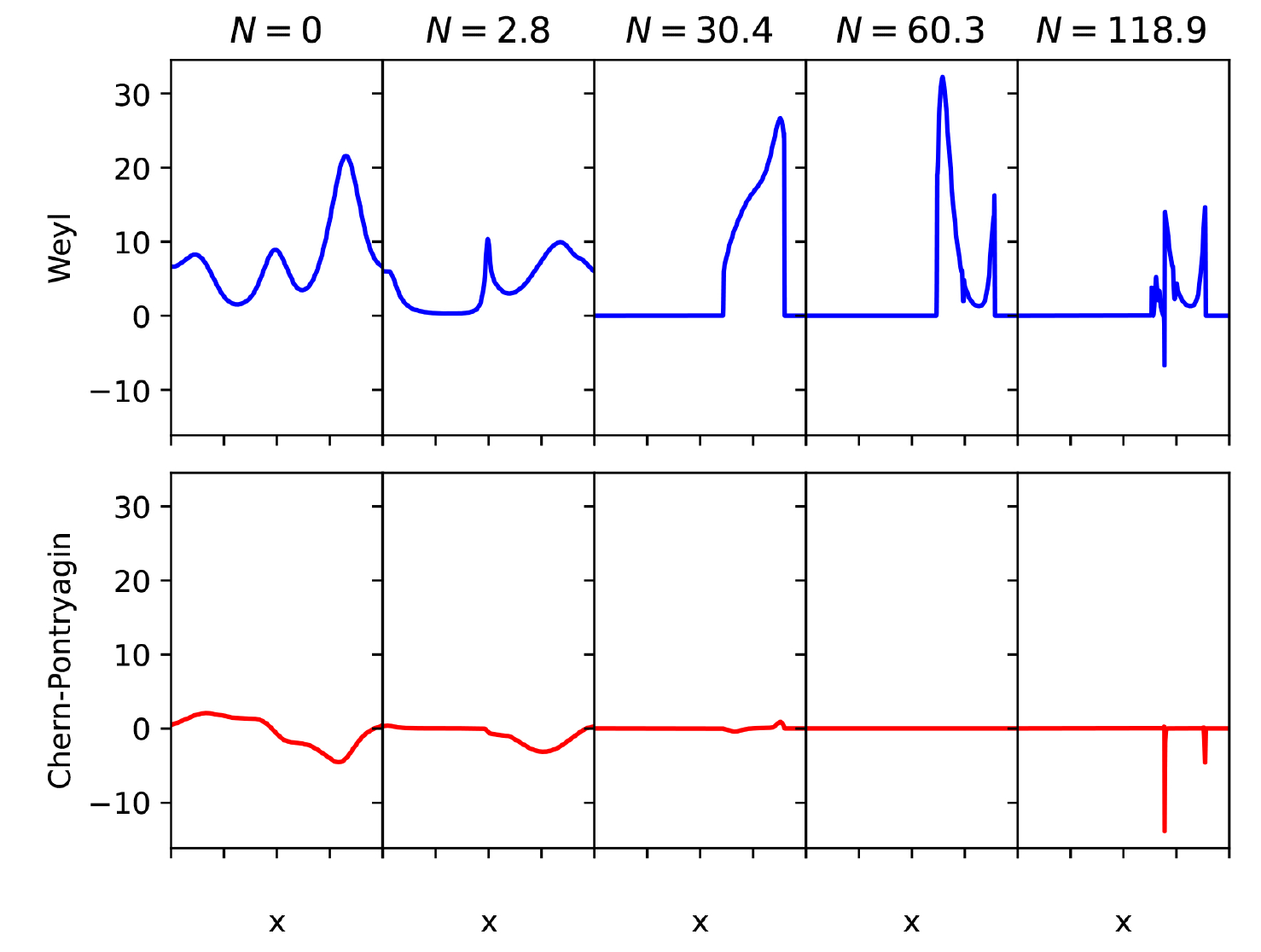}%
    \caption{Snapshots of the Weyl curvature $\bar{\cal C}$ (upper panel) and Chern-Pontryagin invariant  $\bar{\cal P}$ (lower panel) at sample times $N$ during slow contraction corresponding to $M=0.1$ and the initial data set given in Eqs.~(\ref{init2-1}-\ref{init2-4}). The $x$ axis represents the entire simulation domain and the $y$ axis has dimensionless units by construction of the rescaled invariants. The time variable $N$ measures the number of $e$-folds of contraction in the inverse mean curvature $\Theta$. Complete smoothing ($\bar{\cal C}, \bar{\cal P}\lesssim 10^{-10}$) is not reached everywhere. Instead, a small (proper) volume remains in a shear dominated state.}%
    \label{fig:Lim-init}%
\end{figure}

\subsubsection*{Anomalous case}

It is worth comparing this example to another case of slow contraction discussed in Ref.~\cite{Garfinkle:2008ei}, corresponding to a steep(er) negative exponential potential with $M=0.1$ in Eq.~\eqref{def-potential} and
the  free data $\{\Theta_0, \bar{Z}^0_{ab}, \phi, \bar{Q}\}$ as introduced in Sec.~\ref{sec:init_cond}  being specified through the following parameters:
\begin{align}
\label{init2-1}
&\Theta_0 = 3.00;\\
&a_1=0.70,\; a_2=0.10,\; b_1=1.80,\; b_2=-0.15,\; \xi=0.01;
\\
& f_x = 0.15,\; n_x=1,\; h_x =-1.00,\; \phi_0=0.00, ;\\
\label{init2-4}
& q_x =2.00,\; m_x=2,\; d_x=-1.70,\; Q_0 = 0.00.
\end{align}
As shown in Refs.~\cite{Cook:2020oaj,Ijjas:2021wml}, this second example represents a special case that is {\it not} a characteristic outcome of slow contraction. Here, the initial data is tuned such that the spatial 3-curvature $\bar{n}_{ab}, \bar{A}_b$ must remain small throughout the evolution almost everywhere. This leads to an outcome where almost all (proper) volume rapidly approaches the flat FRW geometry but a small (proper) volume  remains in the shear dominated (Kasner-like) region. 

The tuning of the initial data and the resulting evolution during 120 $e$-folds of contraction are reflected in the snapshots of Fig.~\ref{fig:Lim-init}. By Eqs.~(\ref{def-elec}-\ref{def-magn}) and (\ref{def2-weyl}-\ref{def2-cp}), we can see that negligible spatial 3-curvature translates into both the magnetic Weyl tensor and the Chern-Pontryagin invariant  being negligible almost everywhere throughout the evolution, {\it i.e.}, 
$\bar{H}_{ab}\ll \bar{E}_{ab}$ and $\bar{\cal P}\ll\bar{\cal C}$.
By $N\simeq30$, most of spacetime settles into a flat FRW state ($\bar{\cal P},\bar{\cal C}\lesssim 10^{-10}$) but a small (proper volume) region remains in a shear dominated (Kasner-like) state for all times. The shear domination results in a large value of the Weyl curvature in this region:  
\begin{equation}
\bar{\cal C} \simeq 8\bar{E}_{ab} \bar{E}^{ab} \simeq (\bar{\Sigma}_{ab} - \bar{\Sigma}_{\langle a}{}^c\bar{\Sigma}_{ b\rangle c}) (\bar{\Sigma}^{ab} - \bar{\Sigma}^{c\langle a}\bar{\Sigma}_c{}^{b\rangle})\sim{\cal O}(10).
\end{equation}

\begin{figure}[!tb]%
    \centering
\includegraphics[width=0.75\linewidth]{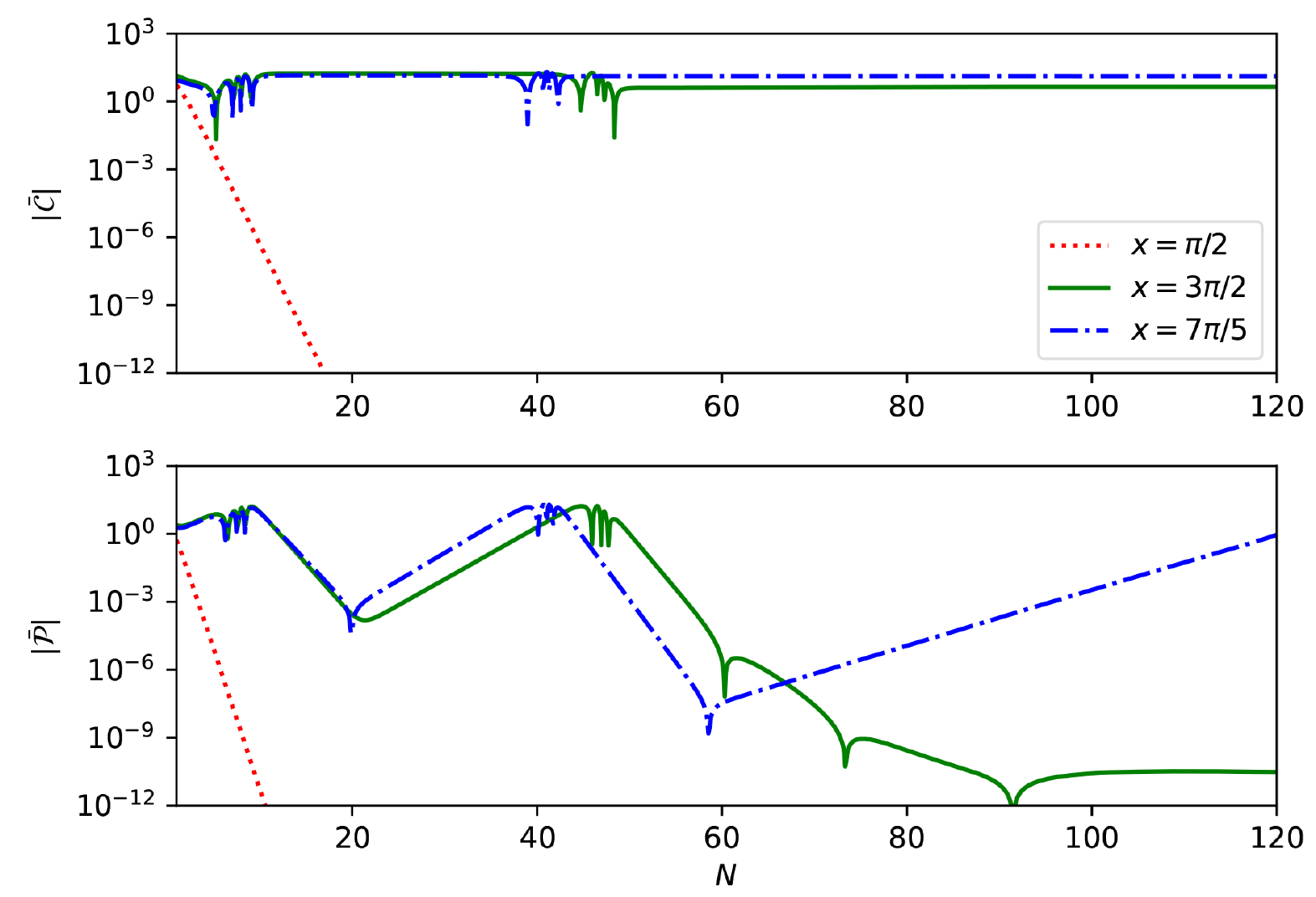}%
    \caption{(Log-scale) Evolution of the Weyl curvature $\bar{\cal C}(x)$ (upper image) and Chern-Pontryagin invariant $\bar{\cal P}(x)$ (lower image) at select spatial points: $x=\pi/2$ (red dotted line),  $x=3\pi/2$ (green solid line) and  $x=16\pi/9$ (blue dash-dotted line) as a function of $e$-fold time $N$ corresponding to the case in Fig.~\ref{fig:Lim-init}.}%
    \label{fig:Lim-numpts_log}%
\end{figure}
The striking difference of the shear dominated region to the large (proper volume) region that quickly settles in the flat FRW state becomes even clearer when comparing the evolution of the curvature invariants at select spatial points in each of the two regions. In Fig.~\ref{fig:Lim-numpts_log}, we depicted the evolution of the Weyl curvature and the Chern-Pontryagin invariant at three distinct spacetime points. The first point ($x=\pi/2$) is in the region that reaches the flat FRW state within a few $e$-folds of contraction. In accordance, the invariants $\bar{\cal C}(x)$ and $\bar{\cal P}(x)$ (red dotted line) rapidly decrease and dip below $10^{-10}$ by $N\simeq17$ exhibiting the qualitatively same behavior as the spacetimes points illustrated in Fig.~\ref{fig:smg-numpts_log}. 
The curvature invariants corresponding to other two points $x=3\pi/2$ (green solid line) and $x=7\pi/5$ (blue dash-dotted line) show rather different behaviors. Both points have in common that the corresponding Weyl curvature starts out at a large value of ${\cal O}(10)$ and -- with the exception of a few `jumps' at $N\sim10$ and $N\sim40$ -- remains nearly the same, albeit having different values at the two points. Despite the similarity that the evolution of  $\bar{\cal C}(x)$ shows at both points and despite the proximity of the two points, the corresponding Chern-Pontryagin invariant behaves surprisingly differently at the two points: initially, $\bar{\cal P}(x)$ undergoes a sequence of growth and shrinkage at both points but, after about 40 $e$-folds, at $x=3\pi/2$, it gradually decreases and settles at a value $\sim 10^{-10}$. At $x=7\pi/5$, on the other hand, the sequence continues with $\bar{\cal P}(7\pi/5)$ approaching $\sim 10$ at $N=120$. Clearly, the local growth of $\bar{\cal P}$ reflects the fact pointed out in Ref.~\cite{Ijjas:2020dws} that, at select spatial points, the 3-curvature components $n_{ab}$ remain non-negligible, driving the local evolution from one shear-dominated (Kasner-like) geometry to another. Capturing this subtle feature makes it clear that the gauge/frame independent diagnostics presented here preserve all nuances of the underlying frame/gauge dependent simulation.

\subsection{Eigenvalue evolution}

Following the evolution of the two curvature invariants for the entire simulation domain as well as at sample spatial points, we  obtain all the  essential gauge and frame independent information we can extract from the simulations. Yet, especially to gain a deeper understanding of the evolution ({\it e.g.}, what are the main contributing effects/quantities to the observed evolution of $\bar{\cal C}$ and $\bar{\cal P}$), it proves useful to extract further, complementary information from the Weyl tensor and associated quantities. 

To this end, we compute the eigenvalues $\lambda_i$ ($i=1,2,3$) corresponding to the electric and magnetic Weyl tensors, $\bar{E}_{ab}$ and $\bar{H}_{ab}$, for details of the computation, see Appendix~\ref{sec:appendixEV}.
Both spatial 3-tensors are real, symmetric and trace-free, meaning that the corresponding three eigenvalues are real and sum to zero.  Consequently, all information entailed in $\bar{E}_{ab}$ and $\bar{H}_{ab}$ can be extracted by following the evolution of two of the corresponding eigenvalues.

%
\begin{figure}[!tb]%
    \centering
\includegraphics[width=0.75\linewidth]{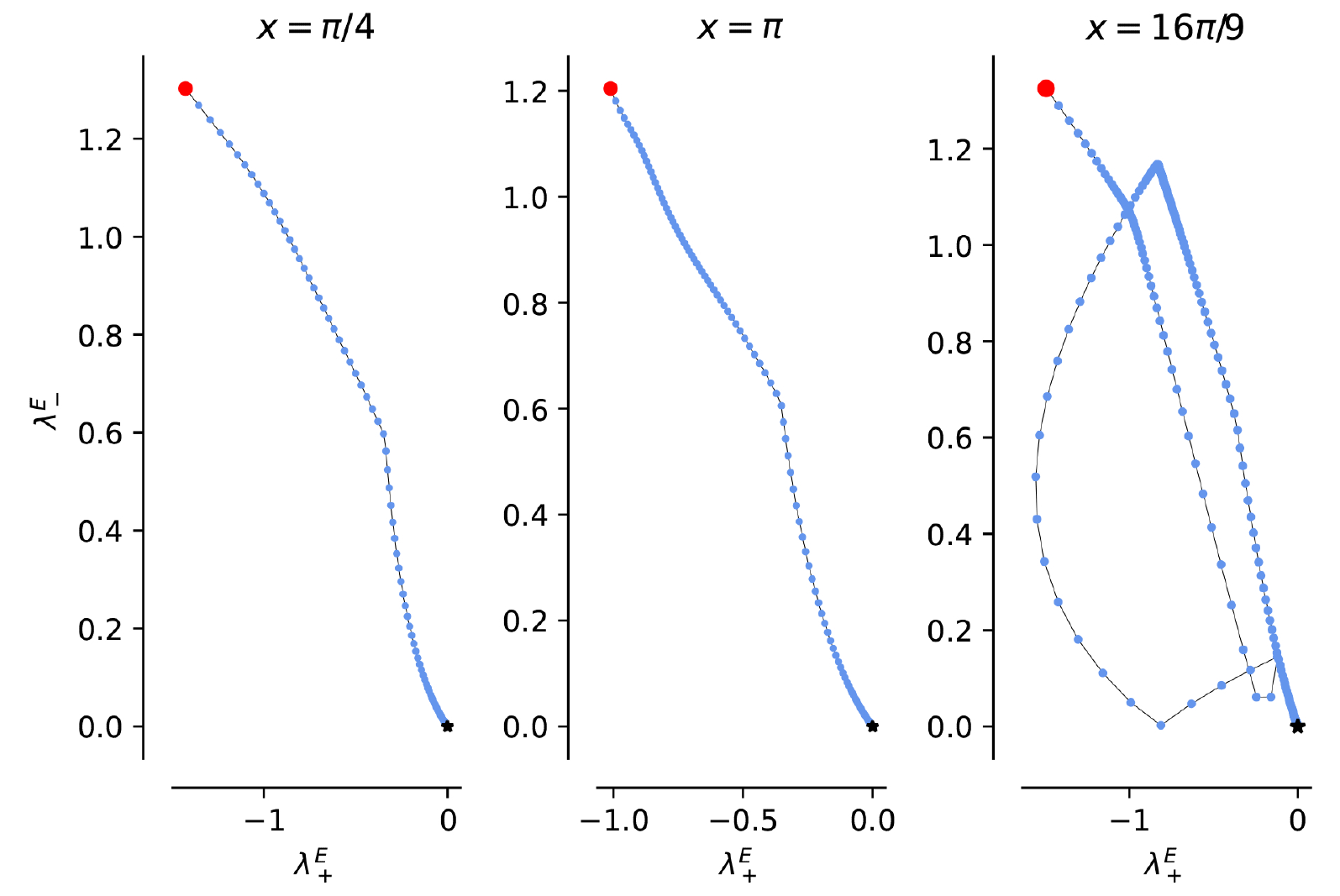}%
    \caption{Evolution of the electric Weyl tensor eigenvalues $\lambda_{\pm}^E$ corresponding to the case in Fig.~\ref{fig:smg-init} at the same select spatial points as in Fig.~\ref{fig:smg-numpts_log}. The red dot denotes $N=0$ and the black star denotes the final state, here $(0,0)$ for each point. The distance in time between two blue dots is $\sim 0.1$ $e$-folds. Darker regions (more points) indicate states in which more time is spent.}%
    \label{fig:smg-Evelec}%
\end{figure}
%
\begin{figure}[!tb]%
    \centering
\includegraphics[width=0.75\linewidth]{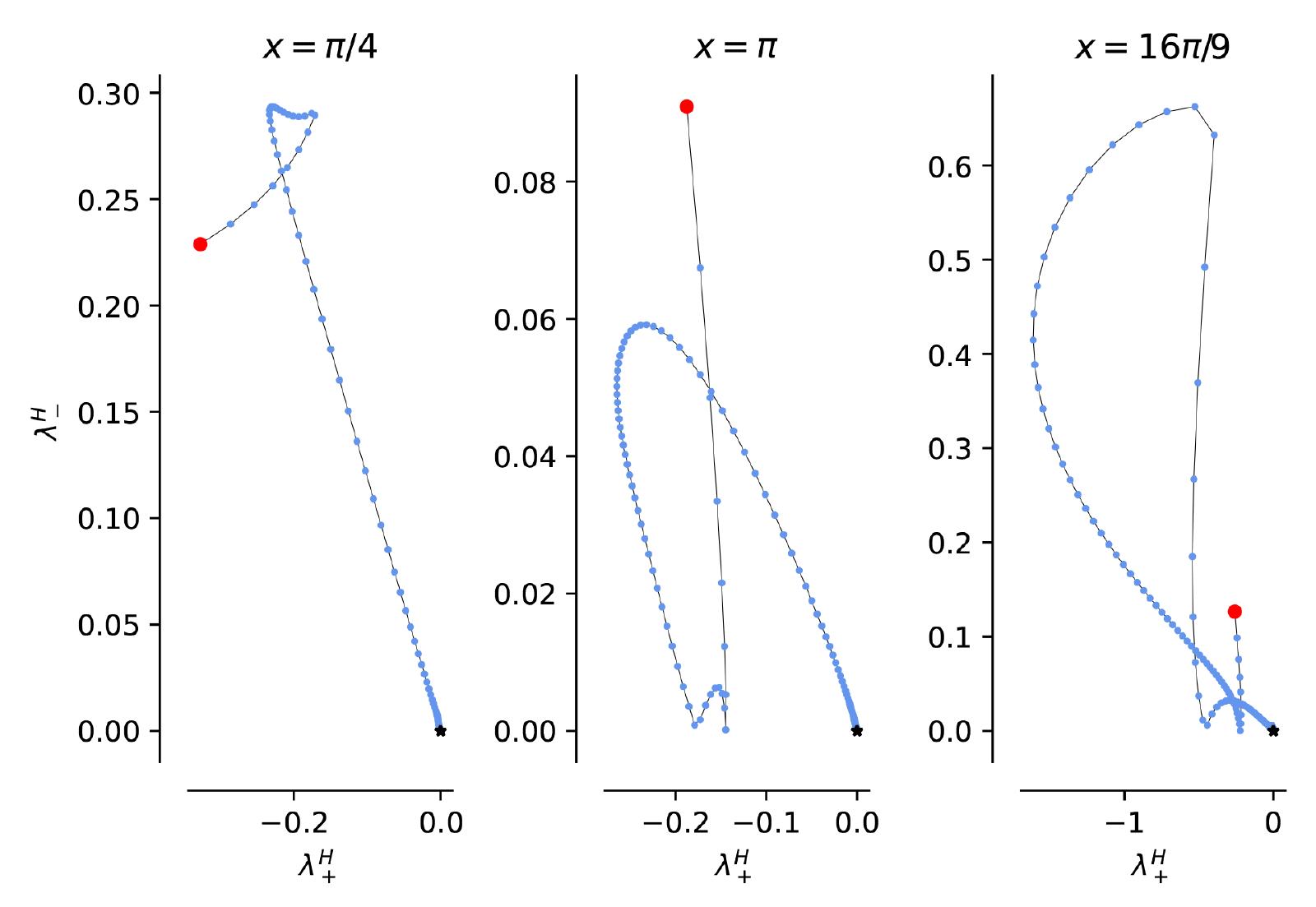}%
    \caption{Evolution of the magnetic Weyl tensor eigenvalues $\lambda_{\pm}^H$ at select spatial points corresponding to the case in Fig.~\ref{fig:smg-init} at the same select spatial points as in Fig.~\ref{fig:smg-numpts_log}. The red dot denotes $N=0$ and the black star denotes the final state, here $(0,0)$ for each point. The distance in time between two blue dots is $\sim 0.1$ $e$-folds. Darker regions (more points) indicate states in which more time is spent.}%
    \label{fig:smg-Evmagn}%
\end{figure}
In Figs.~\ref{fig:smg-Evelec}~and~\ref{fig:smg-Evmagn}, we show orbitspace plots of the eigenvalues 
\begin{eqnarray}
\lambda_{+}^{E,H} &=& {\textstyle \frac12}\Big( \lambda_2^{E,H} + \lambda_3^{E,H} \Big),\\
\lambda_{-}^{E,H} &=&  {\textstyle \frac{1}{2\sqrt{3}}}\Big( \lambda_2^{E,H} - \lambda_3^{E,H} \Big)\\
\end{eqnarray}
corresponding to our main example and the same three spacetime points as in Fig.~\ref{fig:smg-numpts_log}.
Here, $\lambda_{2,3}$ are computed as described in Eqs.~(\ref{eig2}-\ref{eig3}). In each of the plots, the red dot denotes the initial state at $N=0$ and the black star denotes the final state. The fact that all three points settle at $(0,0)$ indicates that they each reach the flat FRW attractor state. The clear difference in the three wordlines is in agreement  with our observation in Figs.~\ref{fig:smg-init}~and~\ref{fig:smg-numpts_log}: different regions settle in the final flat FRW state through different trajectories -- further evidence of ultralocal behavior. 

The blue dots each mark the eigenvalue at a given time. The  time interval $\Delta N$ between two dots is chosen to be $\sim0.1$ $e$-folds of contraction in the inverse mean curvature. (The solid black line is drawn to help guide the eye follow the worldlines.) This way of representing the eigenvalue evolution enables us to read off what time is spent in which state. This is the key added information the state space orbits provide.  In this example, it is clear that the spacetime points spend most of their time at a few select states (darker shaded regions with many points) as opposed to spending equal time at all states they explore. The transitions to these states appear like `jumps.'

Other important information we can read off from the state space orbit  plots is the strong non-linearity of the evolution.The non-linearity is indicated both by the jump-like transitions from one state to another and by the complex shape of the wordlines.

\begin{figure}[tb]%
    \centering
\includegraphics[width=0.85\linewidth]{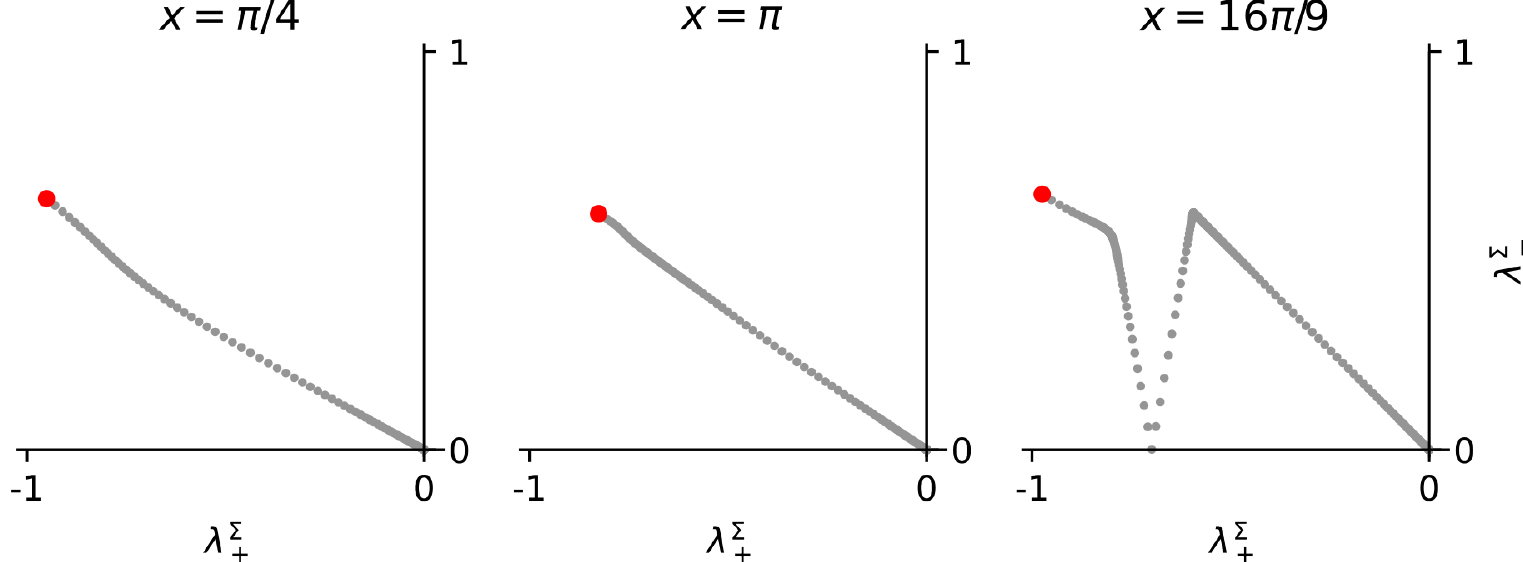}%
    \caption{Evolution of the shear eigenvalues $\lambda_{\pm}^\Sigma$ corresponding to the case in Fig.~\ref{fig:smg-init} at the same select spatial points as in Fig.~\ref{fig:smg-numpts_log}. The red dot denotes $N=0$ and the distance in time between two grey dots is $\sim 0.1$ $e$-folds. Darker regions (more points) indicate states in which more time is spent.}%
    \label{fig:smg-Evsig}%
\end{figure}
\begin{figure}[tb]%
    \centering
\includegraphics[width=0.85\linewidth]{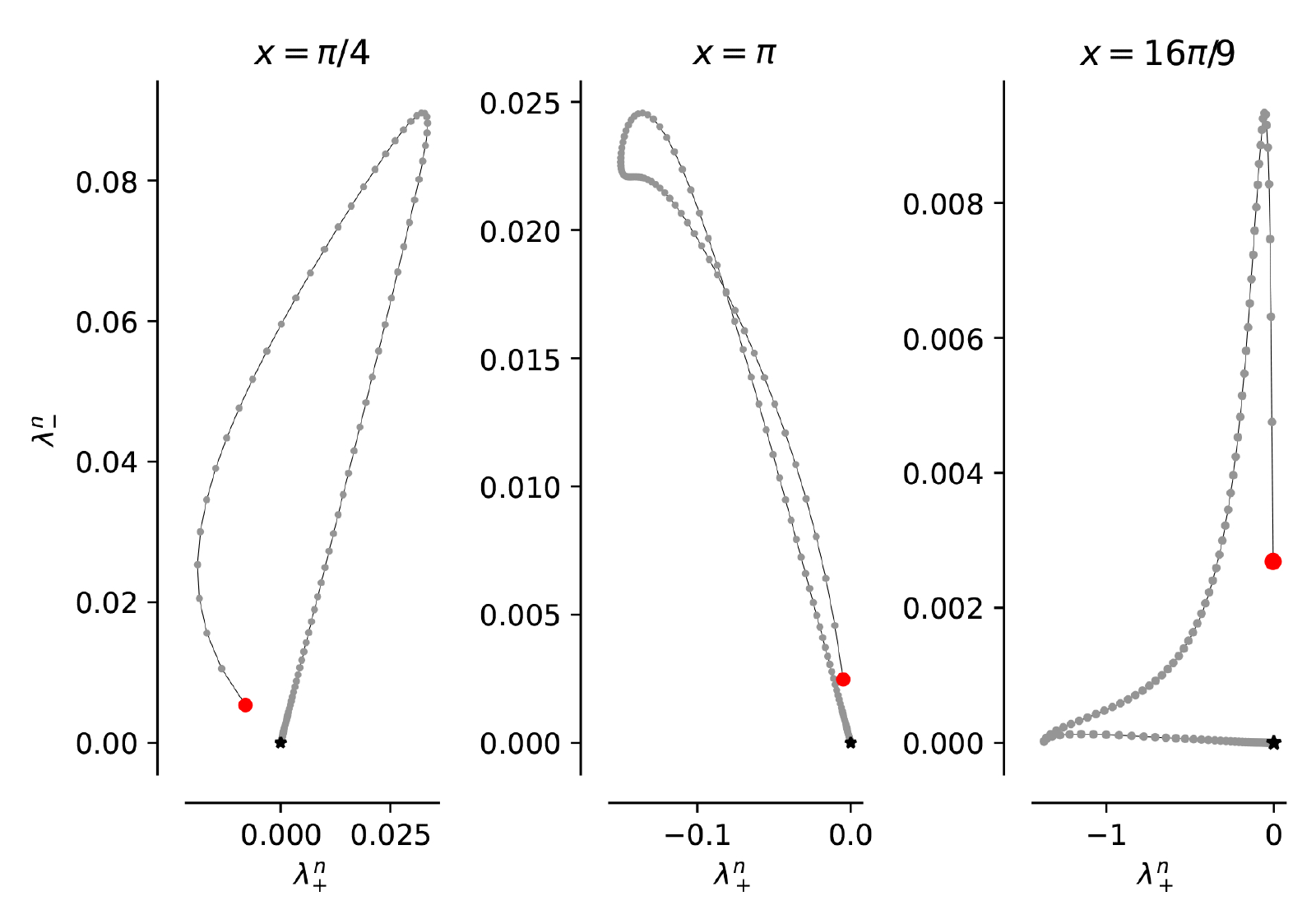}%
    \caption{Evolution of the spatial curvature eigenvalues $\lambda_{\pm}^n$ at select spatial points. The red dot denotes $N=0$ and the distance in time between two grey dots is $\sim 0.1$ $e$-folds. Darker regions (more points) indicate states in which more time is spent.}%
    \label{fig:smg-Evn}%
\end{figure}

Finally, it is worth comparing the evolution of $\lambda_{\pm}^{E}$ and $\lambda_{\pm}^H$ to the eigenvalue evolution of the associated frame dependent quantities, the shear and spatial curvature 3-tensors, $\bar{\Sigma}_{ab}$ and $\bar{n}_{ab}$, respectively. In Figs.~\ref{fig:smg-Evsig}~and~\ref{fig:smg-Evn}, we show state space orbits plots of  $\lambda_{\pm}^{\Sigma}$ and $\lambda_{\pm}^n$. 

The two sets of plots show both similarities and differences to the worldines associated with $\bar{E}_{ab}$ and $\bar{H}_{ab}$. Obviously, all spacetime points reach the point $(0,0)$, in agreement with the flat FRW state being the attractor state. The fact that the transition between different state is jumplike is immediately apparent as well. Less clear is the non-linearity of the evolution: the wordlines have a significantly simpler structure than the worldlines corresponding to the eigenvalues of $\bar{E}_{ab}$ and $\bar{H}_{ab}$. 
Given only the shear state space orbits (as is commonly presented in the relativity literature), one could incorrectly conclude that most of the evolution is equally well-described by perturbative expansion around the homogenous (FRW or Kasner-like) exact solution. However, this is generally not the case, as made evident by the evolution of the curvature invariants.

The main added information we can learn from the worldlines in the $\lambda_{\pm}^{\Sigma}$ and $\lambda_{\pm}^n$ state space orbit plots is which variables (if any) contribute mainly to the evolution of the electric and magnetic Weyl tensor and/or the Weyl curvature and Chern-Pontryagin invariants. Simply comparing the two sets of state space orbit plots is not sufficient for this purposes. But, when running a large set of simulations, it can be very instructive to simultaneously follow the evolution of the eigenvalues corresponding to the electric and magnetic Weyl tensor as well as to the gauge/frame dependent quantities. Here, for example, by using the eigenvalue analysis and running a large number of simulations (involving several hundreds of distinct initial data sets), we could observe that the main contribution to $\bar{\cal C}$ comes from $\bar{E}_{ab}$ while the main contribution to $\bar{E}_{ab}$ comes from  $\bar{\Sigma}_{ab}$. Similarly, we found that the main contribution to $\bar{\cal P}$ comes from $\bar{H}_{ab}$ while the main contribution to $\bar{H}_{ab}$ comes from  $\bar{n}_{ab}$.
This type of complementary analysis is essential when the goal is to find the generic outcomes of a theory/scenario. 

\section{Discussion}

Frame and gauge invariant diagnostics are essential in numerical relativity due to the fact that the scheme used to evolve the Einstein equations is necessarily gauge and/or frame dependent. Such diagnostic methods have been common practice in the context of gravitational wave physics,  {\it i.e.}, to extract the observable gravitational radiation emitted by compact object mergers as evaluated on asymptotically flat spacetimes. But similar tools have been lacking when studying cosmological spacetimes by means of numerical relativity. This has made the result of these studies difficult to interpret and nearly impossible to compare to similar  studies.

In this paper, we have introduced gauge/frame invariant diagnostic  tools that invariantly characterize numerical relativity studies of cosmological spacetimes. Working with an existing example of simulating slow contraction, we have demonstrated that only two quantities -- the Weyl curvature scalar and the Chern-Pontryagin invariant -- suffice to extract all physically relevant information from the cosmological simulations.
In forthcoming papers \cite{Garfinkle:2023vzf, Ijjas:2023dnb}, we will systematically apply our diagnostic tools to analyze, evaluate and compare numerical relativity studies of inflationary expansion and slow contraction.

\subsection*{Acknowledgements.} Many thanks to Paul J. Steinhardt,  David Garfinkle and Frans Pretorius for very helpful comments and discussions. Thanks also to V. (Slava) Mukhanov and Elias Most 
for conversations regarding the gauge problem. This work is supported by the Simons Foundation grant number 947319.

\newpage

\appendix

\section{Numerical scheme}
\label{app-dynvar}

The numerical scheme is presented in great detail in Refs.~\cite{Ijjas:2020dws,Ijjas:2021gkf}. Here, we do not repeat these details but instead restrict ourselves to providing the minimum information so as to keep the description self-contained, only introducing the dynamical variables of the formulation as well as the gauge and frame choices and the resulting evolution and constraint equations.

The orthonormal tetrad {\it formulation} is a vector-based (3+1) formulation of the field equations. That means, as opposed to coordinates, spacetime points are  represented by a quadruple of orthonormal 4-vectors, the timelike vierbein ${\bf e}_0$ and the associated spatial triad $\{{\bf e}_1,{\bf e}_2,{\bf e}_3\}$, with the local metric being flat everywhere, ${\bf e}_{\alpha}\cdot{\bf e}_{\beta}={\rm diag}(-1,0,0,0)$ with $\cdot$ denoting the inner product of the vierbein.

The geometric variables of the scheme are the 16 tetrad vector components $\{\tensor{E}{_\alpha^\mu}\}$ supplemented by the  24 Ricci rotation coefficients $\gamma_{\alpha\beta\lambda} ={\bf e}_{\alpha}\cdot\nabla_{\lambda}{\bf e}_{\beta}$  that describe how the tetrad is being deformed when moving from one spacetime point to another. 
The variables that describe the stress-energy component are the scalar field $\phi$ and its first time and spatial derivatives, $W$ and $S_a$ ($a=1,2,3$), respectively. 

Six of the Ricci rotation coefficients and four of the tetrad vector components are gauge variables. The former determine the tetrad frame's orientation and rotation while the latter relate directional derivatives along tetrads with partial derivatives along coordinate directions. Here, the tetrad {\it frame} is fixed to be minimal, {\it i.e.}, Fermi-propagated (no non-physical rotations) and hypersurface orthogonal (time-like tetrad is normal to space-like hypersurfaces). The coordinate {\it gauge} is specified to be co-moving (zero shift) with the tetrad. In addition, we fix the lapse $N$ by way of an elliptic equation,
\begin{equation}
\label{lapse-eq-H}
 -\Big(\bar{D}_a - 2\bar{A}_a\Big)\bar{D}^a{\cal N}
+ \Big( \bar{\Sigma}_{ab}\bar{\Sigma}^{ab}
+3  + \bar{W}^2 - \bar{V}(\phi)\Big){\cal N} = 3
,
\end{equation}
such that hypersurfaces of constant time are constant mean curvature (CMC) hypersurfaces, where the mean curvature is denoted by $3\Theta^{-1}$. 
Note that specifying a coordinate gauge to supplement the tetrad frame is optional in analytic calculations but it is necessary when the goal is to obtain a partial differential equation (PDE) system ready for numerical integration.  

This particular combination of frame and gauge choices has several advantages, making the resulting especially well-suited for evolving cosmological spacetimes:
\begin{itemize}
\item[-] The remaining free 15 Ricci rotation coefficients 
$\Theta^{-1}, \Sigma_{ab}, n_{ab}, A_b$ have clear physical meaning, with 
\begin{equation}
\Theta^{-1}\equiv {\textstyle \frac13} K_c{}^c 
\end{equation}
denoting the trace of the extrinsic curvature $K_{ab}$,
\begin{equation}
\Sigma_{ab} = K_{ab} - \Theta^{-1}\tensor{\delta}{_a_b}
\end{equation}
denoting the five components of the symmetric, trace-free shear tensor and 
$n_{ab}$ and $A_b$ denoting the nine components of the symmetric and anti-symmetric part of the intrinsic (or spatial) 3-curvature tensor, respectively.
\item[-] The CMC slicing yields a natural time coordinate given by
\begin{equation}
\label{time-def}
e^{-t} = 3 \Theta^{-1},
\end{equation}
such that $N_{\Theta}\equiv t_0-t$ measures the number of $e$-folds of contraction in $\Theta$ between $t_0$ and $t$. In the homogeneous limit, $\Theta$ is the Hubble radius $|H|^{-1}$ such that  $N_{\Theta}$ measures the $e$-folds of contraction in the Hubble radius. 
\item[-] Since $\Theta^{-1}$ is positive definite, we can use it to rescale all of our simulation variables, making them scale-free as follows:
\begin{eqnarray}
\label{cal-N-def}
N &\rightarrow&{\cal N}\equiv N\times \Theta^{-1}
,\\
\{E_a{}^i, \Sigma_{ab}, n_{ab}, A_b\} &\rightarrow&  \{\bar{E}_a{}^i, \bar{\Sigma}_{ab} , \bar{n}_{ab}, \bar{A}_b \} 
\equiv \{E_a{}^i, \Sigma_{ab}, n_{ab}, A_b\} / \Theta^{-1}  
, \\
\{W,  S_a\} &\rightarrow&  \{\bar{W}, \bar{S}_a  \} 
\equiv \{W, S_a\} / \Theta^{-1}  
,\\
\label{Vbar-def}
V &\rightarrow& \bar{V} \equiv V / \Theta^{-2}.
\end{eqnarray}
The variables in Eqs.~(\ref{cal-N-def}-\ref{Vbar-def}) yield a complete set of manifestly frame and gauge dependent dynamical variables of our simulation that we evolve numerically.
\item[-] Rescaling by the mean curvature $\Theta^{-1}$ together with our time coordinate choice $t$ as given in Eq.~\eqref{time-def} enables us to run the simulation for arbitrary long (finite) time without encountering `blow-ups.' This is because, working with scale-free variables, the simulation is not prone to encountering stiffness issues. In addition, curvature singularities are avoided because they cannot be reached within finite coordinate time.
\end{itemize}

In mean-curvature normalized, orthonormal tetrad form, the Einstein-scalar equations take the following form:
\begin{eqnarray}
 \label{eq-E-ai-H}
\partial_t \bar{E}_a{}^i 
&=&  - \Big({\cal N}-1\Big)\bar{E}_a{}^i 
- {\cal N}\bar{\Sigma}_a{}^c \bar{E}_c{}^i
,\\
\label{eq-Sigma-H}
{\partial}_t \tensor{\bar{\Sigma}}{_a_b} 
&=&- \Big(3{\cal N}-1\Big)\tensor{\bar{\Sigma}}{_a_b} 
 + \tensor{\bar{D}}{_\langle _a}\tensor{\bar{D}}{_b_\rangle} {\cal N} 
+\tensor{\epsilon}{^c^d_\langle_a}\tensor{\bar{n}}{_b_\rangle_d} \tensor{\bar{D}}{_c}{\cal N}
+ \tensor{\bar{A}}{_\langle_a}\tensor{\bar{D}}{_b_\rangle}{\cal N} 
+  {\cal N}\tensor{\bar{S}}{_\langle_a}\tensor{\bar{S}}{_b_\rangle}
 \\
&+& {\cal N}\Big( \tensor{\bar{n}}{_c^c}\tensor{\bar{n}}{_\langle_a_b_\rangle}    
- 2 \tensor{\bar{n}}{_\langle_a^c}\tensor{\bar{n}}{_b_\rangle_c}
+\tensor{\epsilon}{^c^d_\langle_a}\tensor{\bar{D}}{_c} \tensor{\bar{n}}{_b_\rangle_d}\Big)
- {\cal N}\Big( 2\tensor{\epsilon}{^c^d_\langle_a}\tensor{\bar{A}}{_c}\tensor{\bar{n}}{_b_\rangle_d}
+ \tensor{\bar{D}}{_\langle_a}\tensor{\bar{A}}{_b_\rangle}\Big)
,\nonumber\\
\label{eq-N-ab-H}
{\partial}_t \tensor{\bar{n}}{_a_b} 
&=& - \Big({\cal N}-1\Big)\tensor{\bar{n}}{_a_b} 
+{\cal N}\Big(
 2 \tensor{\bar{n}}{_(_a^c} \tensor{\bar{\Sigma}}{_b_)_c} 
- \tensor{\epsilon}{_(_a^c^d}\tensor{\bar{D}}{_c} \tensor{\bar{\Sigma}}{_b_)_d} 
\Big)
- \tensor{\epsilon}{^c^d_(_a}\tensor{\bar{\Sigma}}{_b_)_c}\tensor{\bar{D}}{_d}{\cal N} 
,\\
\label{eq-A-b-H}
{\partial}_t \tensor{\bar{A}}{_c} 
&=& - \Big({\cal N}-1\Big)\tensor{\bar{A}}{_c} 
- {\cal N}\Big( \tensor{\bar{\Sigma}}{_c^b}\tensor{\bar{A}}{_b}
-{\textstyle \frac12}\tensor{\bar{D}}{_a} \tensor{\bar{\Sigma}}{_c^a} 
\Big)
-\tensor{\bar{D}}{_c}{\cal N}
+ {\textstyle \frac12}\tensor{\bar{\Sigma}}{^a_c}\tensor{\bar{D}}{_a}{\cal N}
,\\
\label{eq-phi-H}
{\partial}_t \phi &=& {\cal N}\bar{W}
,\\
\label{eq-W-H}
{\partial}_t \bar{W} 
&=& - \Big(3{\cal N}-1\Big)\bar{W}  - {\cal N}\Big( \bar{V},_{\phi}- \bar{D}_a \bar{S}^a  + 2\bar{A}_a \bar{S}^a \Big) + \bar{S}^a\bar{D}_a{\cal N}
,\\
\label{eq-S-a-H}
{\partial}_t \bar{S}_a 
&=& \bar{S}_a -{\cal N}\big(\bar{S}_a + \tensor{\bar{\Sigma}}{_a^b}\bar{S}_b  - \bar{D}_a \bar{W}\Big) + \bar{W}\bar{D}_a{\cal N}  
.
\end{eqnarray}
The evolution system that we solve numerically is subject to following set of constraint equations:
\begin{align}
\label{hamiltonian-const}
 2\bar{D}_b \bar{A}^b &= -3
 +{\textstyle \frac12}\tensor{\bar{\Sigma}}{^a^b}\tensor{\bar{\Sigma}}{_a_b} 
 + {\textstyle \frac12} \tensor{\bar{n}}{^a^b}\tensor{\bar{n}}{_a_b} 
  - {\textstyle \frac14}  (\tensor{\bar{n}}{_a^a})^2 
  +3 \tensor{\bar{A}}{^b}\tensor{\bar{A}}{_b}
 + {\textstyle \frac12}\bar{W}^2 + {\textstyle \frac12}\bar{S}^a\bar{S}_a + \bar{V}(\phi)
,\\
\label{m-const}
\tensor{D}{_b} \tensor{\bar{\Sigma}}{_a^b}  &= 
3\tensor{\bar{\Sigma}}{_a^b}\tensor{\bar{A}}{_b}
+ \tensor{\epsilon}{_a^b^c}\tensor{\bar{n}}{_b^d}\tensor{\bar{\Sigma}}{_c_d}
+ \bar{W} \bar{S}_a
,\\
\label{N-ab-const}
\bar{D}_b \tensor{\bar{n}}{^b_a} &= - \tensor{\epsilon}{^b^c_a}\tensor{\bar{D}}{_b} \tensor{\bar{A}}{_c} + 2\tensor{\bar{A}}{_b}\tensor{\bar{n}}{^b_a}
,\\
\label{const-E-ai-hn}
\tensor{\epsilon}{^b^c_a}\tensor{\bar{D}}{_b} \tensor{\bar{E}}{_c^i} &= \tensor{\epsilon}{^b^c_a} \tensor{\bar{A}}{_b} \tensor{\bar{E}}{_c^i} 
+ \tensor{\bar{n}}{_a^d} \tensor{\bar{E}}{_d^i},
\\
\label{const-phi-H}
\bar{D}_a \phi &= \bar{S}_a.
\end{align}
We do not evolve the constraints. However, they are used to specify constraint satisfying initial data. We also employ them to verify convergence of our code.

\section{Algorithm for computing the eigenvalues of 3x3 symmetric matrices}
\label{sec:appendixEV}

Here, we describe the simple algorithm that we use to numerically compute the eigenvalues corresponding to the spatial 3-tensors $\bar{\Sigma}_{ab}, \bar{n}_{ab}, \bar{E}_{ab}$ and $\bar{H}_{ab}$. 

 Consider a symmetric $3x3$ matrix $M$ of the form
 \begin{equation}
 M\equiv
 \begin{pmatrix}
 m_{11} & m_{12} & m_{13}\\
 m_{12} & m_{22} & m_{23}\\
 m_{13} & m_{23} & m_{33}\\
 \end{pmatrix},
 \end{equation}
 with all $m_{ij} \in \mathbb{R}, (i,j=1,2,3)$.
 The characteristic polynomial of M is given by
  \begin{equation}
  \label{char_M}
 {\rm char}_M(\lambda) = -\lambda^3 + {\rm tr}(M)\lambda^2  
 + \frac12 \Big({\rm tr}(M^2) -{\rm tr}(M)^2\Big)\lambda + {\rm det}(M),
 \end{equation}
 where 
  \begin{eqnarray}
{\rm tr}(M) &\equiv&  m_{11} + m_{22} + m_{33}, \\
{\rm det}(M) &\equiv& m_{11} m_{22} m_{33} 
+ 2 m_{12} m_{13} m_{23}
- m_{11}m_{23}^2  - m_{22} m_{13}^2 - m_{33} m_{12}^2 
, 
\end{eqnarray}
  are the trace and determinant of $M$, respectively, and
  \begin{equation}
{\rm tr}(M^2) \equiv m_{11}^2 + m_{22}^2 + m_{33}^2
+ 2 \left(m_{12}^2 + m_{13}^2  + m_{23}^2 \right)
\end{equation}
is the trace of $M^2$.

For any eigenvalue $\lambda_i$ of $M$,   
\begin{equation}
\label{char_eq}
 {\rm char}_M(\lambda_i) = 0.
\end{equation}
  
If ${\rm det}(M) = 0$, one of the three eigenvalues is zero and the other two eigenvalues $\lambda_{\pm}$, are the zero points of the quadratic polynomial $\lambda_{\pm}^2 - {\rm tr}(M)\lambda_{\pm}  
 - \frac12 \left({\rm tr}(M^2) -{\rm tr}(M)^2\right) =0$,
 \begin{equation}
\lambda_{\pm} = \frac12\left({\rm tr}(M)\pm \sqrt{2\,{\rm tr}(M^2) -{\rm tr}(M)^2}\, \right).
\end{equation}
  
If ${\rm det}(M) \neq 0$, $M$ has three non-zero, real eigenvalues. To compute the eigenvalues of $M$, we use Cardano's method, {\it i.e.}, we introduce the new variable $p$ given by
\begin{equation}
\label{def_t}   
\lambda = -\frac13 \Big(qp -  {\rm tr}(M)\Big), 
\end{equation}
with
\begin{equation}
\label{def_q}
q \equiv   \sqrt{{\rm tr}(M)^2 +  {\textstyle \frac32} \Big({\rm tr}(M^2) -{\rm tr}(M)^2\Big)}
.
\end{equation}
Note that the expression under the square root in Eq.~\eqref{def_q},
  \begin{eqnarray}
{\rm tr}(M)^2 + {\textstyle \frac32}  \left({\rm tr}(M^2) - {\rm tr}(M)^2\right) &=&
3m_{12}^2+ 3m_{13}^2 + 3m_{23}^2\\
&+&  {\textstyle \frac12}(m_{11}-m_{22})^2  
  +   {\textstyle \frac12}(m_{11}-m_{33})^2
  +  {\textstyle \frac12}(m_{22}-m_{33})^2
  \nonumber\\
  &\geq&0,
  \nonumber
  \end{eqnarray}
 is non-negative for any choice of $m_{ij}$ such that $q$ is always real. 
 
 If $q=0$, $M$ is diagonal and $m_{11}=m_{22}=m_{33}$, which are the three identical eigenvalues of $M$.
 
If $q\neq0$, substituting $p$ in Eq.~\eqref{def_t} into Eq.~\eqref{char_M}, we obtain the simple depressed cubic,
    \begin{equation}
 \label{dep_cubic}   
{\rm char}_M(p) = \frac{q^3}{27}\Big( p^3 - 3p
+ r\Big),
\end{equation}
where the zeroth order term $r$ is given by
\begin{equation}
r = \frac{2\, {\rm tr}(M)^3 + \frac92 {\rm tr}(M) \left({\rm tr}(M^2)-{\rm tr}(M)^2\right) + 27{\rm det}(M)}{q^3}\,.
\end{equation} 
 
For any solution $p$ of ${\rm char}_M(p) = 0$ with $|p|\leq2$, we can substitute $p=2\,\cos\vartheta$ into Eq.~\eqref{dep_cubic} and obtain
 \begin{equation}
{\rm char}_M(p) = 0\quad \Leftrightarrow \quad \vartheta = \frac13 {\rm arccos}\left(\frac{r}{2}\right).
\end{equation}
In this case, we obtain the three eigenvalues of $M$ after :
\begin{eqnarray}
\label{eig1}
\lambda_1 &=& {\textstyle \frac13}\,{\rm tr}(M) 
- {\textstyle \frac23} q \cos \vartheta,
\\
\label{eig2}
\lambda_2 &=& {\textstyle \frac13}\,{\rm tr}(M) 
+ {\textstyle \frac13} q  \left( \cos \vartheta + \sqrt{3} \sin \vartheta \right)
,\\
\label{eig3}
\lambda_3 &=& {\textstyle \frac13}\,{\rm tr}(M) 
+ {\textstyle \frac13} q  \left( \cos \vartheta - \sqrt{3} \sin \vartheta \right).
\end{eqnarray}

Assume there exists a solution $p$ of ${\rm char}_M(p) = 0$ with $|p|>2$, then $|r|>2$. This means that $p$ is complex because the discriminant $\Delta$ of the  corresponding depressed cubic,
\begin{equation}
\Delta = 27(4-r^2) <0.
\end{equation}
However, the eigenvalues of symmetric real matrices are all real, {\it i.e.}, $\Delta$ must be non-negative, or equivalently, $|r|<2$. Hence, there are no solutions of ${\rm char}_M(p) = 0$ with $|p|>2$.

At each time step $t$, we numerically compute the three eigenvalues corresponding to the spatial 3-tensor $\bar{\Sigma}_{ab}(t,x_s), \bar{n}_{ab}(t,x_s), \bar{E}_{ab}(t,x_s)$ and $\bar{H}_{ab}(t,x_s)$ computed at sample spatial points $x_s$ using Eqs.~(\ref{eig1}-\ref{eig3}).

Note that, unlike in exact analytics where for real, symmetric $3x3$ matrices $|r/2|\leq1$, in numerical computations the term $r/2$ can take values slightly below $-1$ or slightly above $1$ due to truncation error. In such cases, the angle $\vartheta$ and hence the eigenvalues returned by the computation would be complex as opposed to real numbers. To avoid this truncation problem, we check if $r/2<-1$ and, in such instances, set $\vartheta=\pi/r$; and we check if $r/2>1$ and, in such instances, set $\vartheta=0$.

\bibliographystyle{plain}
\bibliography{weyl}

\end{document}